\documentclass[english]{emulateapj}
\usepackage{mathptmx}
\usepackage[T1]{fontenc}
\usepackage{verbatim}
\usepackage{graphicx}
\usepackage{amssymb}

\makeatletter

\providecommand{\tabularnewline}{\\}

\usepackage{times}
\DeclareFontFamily{OT1}{pzc}{}
\DeclareFontShape{OT1}{pzc}{m}{it}%
              {<-> s * pzcmi8t}{}
\DeclareMathAlphabet{\mathpzc}{OT1}{pzc}%
                                 {m}{it}

\AtBeginDocument{
  
}

\makeatother

\usepackage{babel}

\begin{document}
\newcommand{\Ms}{M_{\star}}
\newcommand{\Rs}{R_{\star}}
\newcommand{\Ls}{L_{\star}}
\newcommand{\nstr}{n_{\star}}
\newcommand{\Mo}{M_{\odot}}
\newcommand{\Ro}{R_{\odot}}
\newcommand{\Lo}{L_{\odot}}
\newcommand{\Mbh}{M_{\bullet}}
\newcommand{\SgrA}{\mathrm{Sgr}A^{\star}}
\newcommand{\rstr}{r_{\star}}
\newcommand{\T}[1]{\!#1\!}

\title{Strong mass segregation around a massive black hole}

\author{Tal Alexander%
\footnote{Incumbent of the William Z. \& Eda Bess Novick career development
chair%
} }

\affil{Faculty of Physics, Weizmann Institute of Science, P.O. Box 26, Rehovot
76100, Israel}

\email{tal.alexander@weizmann.ac.il }

\and{}

\author{Clovis Hopman }

\affil{Leiden University, Leiden Observatory, P.O. box 9513, NL-2300 RA
Leiden, The Netherlands}

\email{clovis@strw.leidenuniv.nl}

\begin{abstract}
We show that the mass-segregation solution for the steady state distribution
of stars around a massive black hole (MBH) has two branches: the known
weak segregation solution \citep{bah+77}, and a newly discovered
strong segregation solution, presented here. The nature of the solution
depends on the heavy-to-light stellar mass ratio $M_{H}/M_{L}$ and
on the unbound population number ratio $N_{H}/N_{L}$, through the
relaxational coupling parameter $\Delta\!=\!4N_{H}M_{H}^{2}\left/\left[N_{L}M_{L}^{2}(3\!+\! M_{H}/M_{L})\right]\right.$.
When the heavy stars are relatively common ($\Delta\!\gg\!1$), they
scatter frequently on each other. This efficient self-coupling leads
to weak mass segregation, where the stars form $n\!\propto\! r^{-\alpha_{M}}$
mass-dependent cusps near the MBH, with indices $\alpha_{H}\!=\!7/4$
for the heavy stars and $3/2\!<\!\alpha_{L}\!<\!7/4$ for the light
stars (i.e. $\max(\alpha_{H}\!-\!\alpha_{L})\!\simeq\!1/4$). However,
when the heavy stars are relatively rare ($\Delta\!\ll\!1$), they
scatter mostly on light stars, sink to the center by dynamical friction
and settle into a much steeper cusp with $2\!\lesssim\!\alpha_{H}\!<\!11/4$,
while the light stars form a $3/2\!<\!\alpha_{L}\!<\!7/4$ cusp, resulting
in strong segregation (i.e. $\max(\alpha_{H}\!-\!\alpha_{L})\!\simeq\!1$).
We show that the present-day mass function of evolved stellar populations
(coeval or continuously star forming) with a universal initial mass
function, separate into two distinct mass scales, $\sim\!1\,\Mo$
of main sequence and compact dwarfs, and $\sim\!10\,\Mo$ of stellar
black holes (SBHs), and have $\Delta\!<\!0.1$. We conclude that it
is likely that many relaxed galactic nuclei are strongly segregated.
We review indications of strong segregation in observations of the
Galactic Center and in results of numeric simulations, and briefly
list some possible implications of a very high central concentration
of SBHs around a MBH.
\end{abstract}


\section{Introduction}

\label{s:intro}

Early theoretical studies of the dynamics and distribution of stars
around a MBH \citep{pee72,bah+76,bah+77,you80} were triggered by
the discovery of quasars \citep[e.g.][]{mat+63,sch63} and the realization
that many galactic nuclei may contain a central massive collapsed
object \citep{lyn69,wol+70}, %
{} as well as by the discovery of X-ray sources in globular clusters
\citep{gia+72}, which were then thought to be accreting MBHs (e.g.
\citealt{wyl70,bah+76}; see also \citealt{mil+02b} for a recent
reevaluation of this possibility). The main motivations for these
studies were the prospect of detecting MBHs by the observed stellar
density profile and by tidal disruption flares, and the possible role
of tidal disruptions of stars in the growth of MBHs \citep{lig+77,coh+78,sha+78,ree88}

The renewed interest in this problem is driven by observations of
stars closely orbiting the Galactic MBH \citep{eis+05,ghe+05} and
the accumulating data on their distribution and dynamics \citep{ale99,gen+00,gen+03a,sch+03,sch+07},
as well as by the prospects of detecting gravitational waves (GW)
from extreme mass ratio inspiral sources by future GW detectors (EMRIs:
compact remnants inspiraling into MBHs, see review by \citealt{ama+07};
\citealt{hop06}). EMRI rates and properties strongly depend on the
stellar density and the stellar dynamical processes within $O(0.01\,\mathrm{pc})$
of the MBH, where inspiraling sources originate \citep[e.g.][]{hop+05,hop+06a}. 

Mass segregation occurs in dynamically relaxed systems. MBHs are naturally
expected to lie in relaxed cores in scenarios where the MBH is formed
by run-away mergers in the extreme central density following core
collapse \citep{ree84}, which occurs on timescales much longer then
the relaxation time, $T_{R}$\texttt{\textbf{ }}\citep[e.g. ][]{spi87,qui96b,fre+06b,fre+06c}.
Likewise, the extreme mass ratio targets of the planned Laser Interferometer
Space Antenna%
\footnote{See LISA mission website\texttt{ http://lisa.nasa.gov}%
} GW detector (LISA) are expected to originate in relaxed nuclei, since
LISA design is sensitive to GW from MBHs with mass $\lesssim\!10^{7}\, M_{\odot}$.
The observed correlation between the MBH mass $\Mbh$ and the typical
velocity dispersion of the spheroid of the host galaxy, $\Mbh\!\propto\!\sigma^{\beta}$,
$4\!\lesssim\!\beta\!\lesssim5$ \citep[the $\Mbh/\sigma$ relation, ][]{fer+00,geb+00}
then implies that such nuclei are dynamically relaxed and very dense
\citep{ale07,mer+07}. This can be seen by assuming for simplicity
$\beta\!=\!4$ (a higher value only reinforce these conclusions),
and noting that the MBH radius of influence $r_{h}\!\sim\! G\Mbh/\sigma^{2}\!\propto\!\!\Mbh^{1/2}$
encompasses a stellar mass of order $\Mbh$, so that the number of
stars there is $N_{h}\!\sim\! M_{\bullet}/\Ms$, where $\Ms$ is the
typical stellar mass, and the mean stellar density is $\bar{n}_{h}\!\sim\! N_{h}/r_{h}^{3}\!\propto\! M_{\bullet}^{-1/2}$.
The {}``$nv\Sigma$'' rate estimate of strong gravitational collisions
then implies that $T_{R}^{-1}(r_{h})\!\sim\!\bar{n}_{h}\sigma(G\Ms/\sigma^{2})^{2}\!\propto\!\Mbh^{-5/4}$.
A more rigorous estimate shows that for the Galactic MBH ($\Mbh\!\simeq\!4\times10^{6}\,\Mo$,
\citealt{eis+05,ghe+05}), an archetype of LISA targets, $T_{R}\!\sim\! O(1\,\mathrm{Gyr})\!<t_{H}$
(the Hubble time) and $\bar{n}_{h}\!\sim\! O(10^{5}\,\mathrm{pc^{-3}})$.
The density in the stellar cusp near the MBH is orders of magnitude
higher still (see below). Since $T_{R}\!\propto\!\Mbh^{-5/4}$, isolated
nuclei with $\Mbh\!\lesssim\!10^{7}\,\Mo$ are predicted to be relaxed.

A single mass stellar system around a MBH is expected to relax to
a $r^{-\alpha}$ cusp with $\alpha\!=\!7/4$ \citep{bah+76}. This
results from the fact that the gravitational orbital energy gained
by the system when stars are destroyed near the MBH is conserved as
it is shared and carried outward by the remaining stars, $\dot{E}(r)\!\sim\! E(r)N(<r)/T_{R}\propto r^{-1}r^{3-\alpha}/r^{\alpha-3/2}\!=\! r^{7/2-2\alpha}\!=\mathrm{const}$
\citep{bin+87}. When the system includes a spectrum of masses, $M_{L}\le\! M\!\le\! M_{H}$,
the approach toward equipartition by 2-body interactions decreases
the specific kinetic energy of the high-mass stars, while that of
the low-mass stars increases. As a result, the high-mass stars sink
and concentrate in the center on the dynamical friction timescale
$T_{\mathrm{df}}\T{\sim}T_{R}\left\langle M\right\rangle /M_{H}$,
while the low-mass stars float out \citep{spi87}. 

\citet[hereafter BW77]{bah+77} approximated the mass segregation
problem in the Fokker-Planck formalism, and solved for the steady
state, angular momentum averaged stellar distribution functions (DFs)
$f_{M}(E)$. They found that near the MBH, the DFs can be approximated
by power-laws, $f_{M}(E)\!\propto\! E^{p_{M}}$, and that the stellar
current into the MBH, $Q_{M}(E)$, is very small, which leads to a
specific relation between the stellar mass and the logarithmic slope
of the DF, $p_{M}\!=\! M/4M_{H}$. In the Keplerian limit near the
MBH, these DFs correspond to power-law density cusps, with $\alpha_{M}\!=\!3/2+p_{M}$.
The BW77 solution thus predicts a relatively small range of central
concentrations, from $\alpha_{H}\!=\!7/4$ for the heaviest stars
in the populations, to $\alpha_{L}\!\rightarrow\!3/2$ for the lightest
stars. We show below that the {}``zero-flow'' assumption breaks
down when the massive stars are too rare to scatter efficiently against
each other, and instead sink to the center by dynamical friction against
the light stars. As a result, the BW77 relation between $M$ and $p_{M}$
no longer holds, and the range of central concentrations far exceeds
that of the BW77 solution.

The Galactic Center (GC) provides to date the few available observations
that directly bear on the question of mass segregation around a MBH.
The over-abundance of X-ray transients in the central $\sim\!1$ pc
of the GC was interpreted as evidence of a high central concentration
of neutron stars and stellar black holes (SBHs) \citep{mun+04}; The
central decrease in the surface density of the low-mass horizontal
branch {}``red clump'' giants was interpreted as evidence of the
evacuation of long-lived light objects by mass segregation \citep{lev06b,sch+07}.
The dynamical upper limit on the distributed dark mass in the inner
few mpc around the galactic MBH, $M_{\mathrm{dm}}/\Mbh\!\sim\!\mathrm{few\times}0.01$
\citep{mou+05,gil+08,ghe+08}, is still 10--100 times larger than
predicted by approximate theoretical estimates \citep{mor93,mir+00},
Fokker-Planck calculations \citep{hop+06b}, or by the conservative
drain limit \citep{ale+04}, which places an upper bound on the maximal
number of compact remnants that can avoid being thrown into the MBH
by mutual 2-body scattering. These theoretical estimates are consistent
with the upper bound derived from the observed limits on the diffuse
X-ray in the central 0.7 pc ($\sim2\times10^{4}$) \citep{dee+07}.

Detailed analysis of observations of the GC \citep{lev06b,sch+07}
suggests that the degree of segregation between the light and heavy
stars is stronger than expected in the BW77 solution, although this
interpretation of the data is not unique. Strong segregation in the
GC is further supported by analytic \citep{hop+06b} and Monte-Carlo
\citep[M. Freitag, priv. comm.]{fre+06} mass segregation results
(Fig. \ref{f:FPGCmodel}). These calculations do not take into account
star formation and evolution, but instead assume for simplicity a
non-evolving mass function that is based on a model of the GC's present
day mass mass function (PMF) as an old, continuously star forming
population with the {}``universal'' initial mass function (IMF)
\citep[table 2.1]{ale+99a,ale05}. In these theoretical models $\alpha_{L}\!\simeq\!1.5$,
while $\alpha_{H}\!\gtrsim\!2$ at $r\!\sim\!0.1$ pc. $N$-body simulations
of stellar clusters with evolving stellar populations and a central
intermediate mass black hole (IMBH) also provide hints of strong mass
segregation \citep[J. Makino, priv. comm.]{bau+04}. As we show below
(Fig. \ref{f:pLH_VR}), these results reflect the fact that the relative
fraction of SBHs in old populations is below a critical threshold
needed for them to scatter efficiently against each other and maintain
the weak segregation solution. Instead, they sink to the center by
dynamical friction and settle into the strong segregation solution. 

This paper is organized as follows. The Fokker-Planck formulation
of the mass segregation problem, the choices of boundary conditions
and the various simplifying approximations are described in \S \ref{s:FPeq}.
The physical meaning of the relaxational coupling parameter, $\Delta$,
is discussed in \S \ref{s:rlxparm}. The $\Delta$ parameters of
different stellar populations are explored in \S \ref{s:PMF}. A
large grid of Fokker-Planck mass segregation calculations is presented
and compared to simple analytic estimates in \S \ref{s:results}.
The results and their implications are discussed and summarized in
\S \ref{s:discuss}.

\section{The Energy diffusion Fokker-Planck equation}

\label{s:FPeq}

The MBH dominates the gravitational potential within its radius of
influence, where the dynamics are approximately Keplerian. Following
the approach introduced by \citet{bah+76,bah+77} and its implementation
by \citet{hop+06b}, we calculate the approximate steady state DFs
of a non-evolving stellar population in a spherical Keplerian system
around a fixed MBH. This is done by solving the time and energy-dependent,
angular momentum-averaged particle conservation equation that describes
the 2-body diffusion of stars in energy from a fixed unbound reservoir
into the MBH sink (resonant relaxation, \citealt{rau+96}, is not
taken into account). One of the stellar mass groups, of mass $\Ms$,
is taken to be the reference star. It is assumed that unbound stars
of mass $M$ with velocity dispersion $\sigma_{M}^{2}$ and specific
energy $\epsilon\equiv\psi-v^{2}/2\!<\!0$ ($\psi=-\phi$ is the negative
of the gravitational potential $\phi$ and $\epsilon>0$ for bound
stars), have a Maxwellian DF, 

\begin{equation}
f_{M}(\epsilon)=S_{M}\frac{\nstr}{(2\pi\sigma_{M}^{2})^{3/2}}\exp(\epsilon/\sigma_{M}^{2})\,,\qquad(\epsilon<0)\,,\label{e:MB}\end{equation}
where $n_{\star}$ is the number density normalization of the reference
star. Since the number density of unbound stars \begin{eqnarray}
n_{M} & = & 4\pi\int_{\sqrt{2\psi}}^{\infty}v^{2}f_{M}(\epsilon)\mathrm{d}v=4\pi\int_{-\infty}^{0}\sqrt{2(\psi-\epsilon)}f_{M}(\epsilon)\mathrm{d}\epsilon\nonumber \\
 & = & S_{M}\nstr\left[\frac{2}{\sqrt{\pi}}\frac{\sqrt{\psi}}{\sigma_{M}}+\exp\left(\frac{\psi}{\sigma_{M}^{2}}\right)\mathrm{erfc}\left(\frac{\sqrt{\psi}}{\sigma_{M}}\right)\right]\nonumber \\
 & \rightarrow & S_{M}\nstr\qquad(\psi/\sigma_{M}^{2}\rightarrow0)\,,\end{eqnarray}
$ $the quantity $S_{M}\nstr$ can be interpreted as the total density
of stars of mass $M$ far from the MBH, where $\psi/\sigma_{M}^{2}\!\ll\!1$
and almost all stars are unbound to the MBH. In that case $S_{M}$
is the asymptotic number density ratio of star $M$ relative to the
reference star ($S_{\star}\!=\!1$ by definition). The asymptotic
population ratios between stars of mass $M_{H}$ and $M_{L}$ are
designated below by $N_{H}/N_{L}=S_{H}/S_{L}$. It is assumed that
the asymptotic PMF, and its corresponding population ratios $N_{H}/N_{L}$,
depend only on the IMF, on the star formation history and on stellar
evolution physics, and are independent of mass segregation dynamics,
which are effective only near the MBH. 

Stellar dynamics near the MBH are described here in units where $G\!=\!1$,
mass is measured in units of the mass of the reference star, $\Ms$,
specific energy in units of its velocity dispersion $\epsilon_{\star}\!=\!\sigma_{\star}^{2}$,
and time in units of its two-body relaxation time at the radius of
influence, \begin{equation}
t_{\star}=\frac{3(2\pi\sigma_{\star}^{2})^{3/2}}{32\pi^{2}G^{2}M_{\star}^{2}n_{\star}\ln\Lambda}\,,\end{equation}
where the Coulomb term is estimated as $\Lambda\!=\!\Mbh/\Ms$. Phase
space density is expressed in units of $f_{\star}\!=\! n_{\star}/(2\pi\sigma_{\star}^{2})^{3/2}$
and distance in expressed units of the MBH radius of influence $\rstr\!=\! G\Mbh/\sigma_{\star}^{2}$.
In these units, the dimensionless specific orbital energy is defined
as $x\!=\!\epsilon/\sigma_{\star}^{2}\!=\!\rstr/2a$ ($a$ is the
semi-major axis), the dimensionless time is defined as $\tau\!=\! t/t_{\star}$,
and the dimensionless DF as $g_{M}\!=\! f_{M}/f_{\star}$. 

The evolution of the dimensionless DF of each stellar mass group,
$g_{M}$, is described by the energy flow integral $Q_{M}$, which
expresses the change in energy due to 2-body scattering, and the angular
momentum-averaged effective loss-cone term $R_{M}$ \citep{bah+76,bah+77},

\begin{equation}
\frac{\partial}{\partial\tau}g_{M}(x,\tau)=-x^{5/2}\frac{\partial}{\partial x}Q_{M}(x,\tau)-R_{M}(x,\tau)\,.\label{e:FPeq}\end{equation}
The flow integral is given by

\begin{eqnarray}
Q_{M}(x) & = & \sum_{M'}MM'\int_{-\infty}^{x_{D}}\frac{\mathsf{d}x'}{\max\left(x,x'\right)^{3/2}}\times\nonumber \\
 &  & \left[g_{M}(x)\frac{\partial g_{M'}(x')}{\partial x'}\!-\!\frac{M'}{M}g_{M'}(x')\frac{\partial g_{M}(x)}{\partial x}\right]\,.\label{e:Qm}\end{eqnarray}
 The averaged effective loss-cone term in the diffusive regime (defined
here as $x\T{>}x_{\mathrm{diff}}\T{=}10$) is approximated by \citep{hop+06b}\begin{equation}
R_{M}(x)\!=\!\frac{g_{M}(x)}{\tau_{r}(x)\ln[J_{c}(x)/J_{lc}]}\,,\label{e:Rm}\end{equation}
where $J_{c}(x)$ is the maximal (circular) angular momentum for energy
$x$, $J_{lc}$ is the angular momentum of the loss-cone (taken here
to be $J_{LSO}\T{=}4G\Mbh/c$, the angular momentum of the last stable
orbit for $\left|\epsilon\right|\T{\ll}c^{2}$), and $\tau_{r}\!\sim\!\left[{\normalcolor \sum_{M}g_{M}(x)M^{2}}\right]^{-1}$
is the approximate 2-body relaxation time. The contribution of the
loss-cone in the full loss-cone regime, which typically corresponds
to $x\T{\lesssim}1/2$, is small \citep[e.g.][]{sye+99}, and is therefore
neglected here. \citet{bah+76,bah+77} argued that the loss-cone term
has only a small effect on the energy dependence of the DF, and neglected
it in their multi-mass models. In order to allow direct comparison
of our calculations with their results, we omit here the loss-cone
term as well, unless stated otherwise. 

\begin{figure*}
\noindent \begin{centering}
\begin{tabular}{cc}
\includegraphics[width=1\columnwidth]{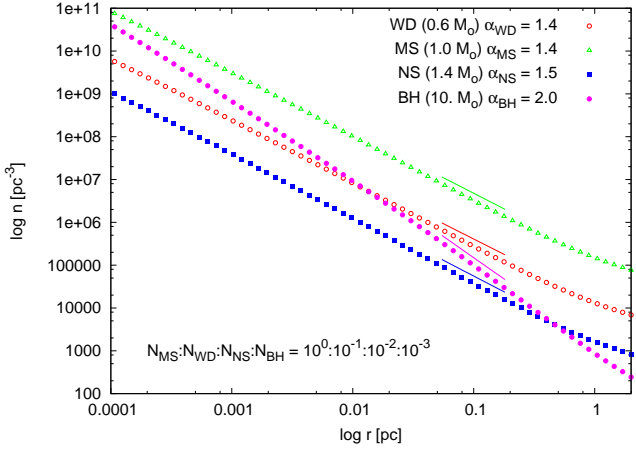} & \includegraphics[width=1\columnwidth]{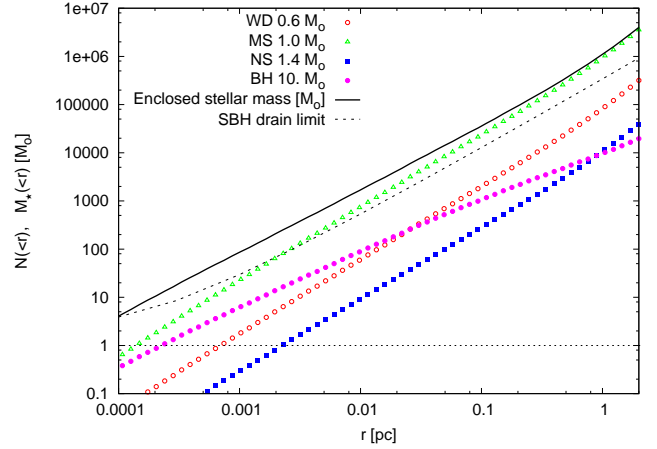}\tabularnewline
\end{tabular}
\par\end{centering}

\caption{\label{f:FPGCmodel} A Fokker-Planck model of the GC assuming a non-evolving
PMF based on the continuous star-forming model shown in Fig. (\ref{f:PMFdelta}),
with violent relaxation boundary conditions, a loss-cone term \citep{hop+06b}
and assuming $\Mbh\!=\!4\times10^{6}\,\Mo$, $\Ms\!=\!1\,\Mo$, $\rstr\!=\!2\,\mathrm{pc}$,
and a total stellar mass $\Mbh$ within $\rstr$. Left: The spatial
number density of the 4 mass components (Eq. \ref{e:nM}). Also shown
are the best-fit logarithmic slopes around $r\!=\!0.1$ pc. Right:
The corresponding enclosed stellar number and total stellar mass as
function of radius. The SBH number is consistent with the drain limit
\citep{ale+04}, also shown. The enclosed number of SBHs within 0.1
pc (0.01 pc) is approximately $1.2\times10^{3}$ ($1.1\times10^{2}$),
of neutron stars $3.2\times10^{2}$ ($12$), of main sequence stars
$2.7\times10^{4}$ ($9.4\times10^{2}$) and of white dwarfs $2.3\times10^{3}$
($76$).}

\end{figure*}

Equation (\ref{e:FPeq}) is integrated in time from an arbitrary initial
DF until steady-state is achieved, subject to the boundary conditions
that the DF falls to zero at some very high energy $x_{D}$ where
the stars are destroyed, and that the unbound stars are replenished
from a Maxwellian reservoir,

\begin{equation}
g_{M}(x\!>\! x_{D})\!=\!0\,,\quad g_{M}(x\!<\!0)\!=\! C_{M}\exp[(\sigma_{\star}^{2}/\sigma_{M}^{2})x]\,,\label{e:BCs}\end{equation}
 where the constant $C_{M}$ is related to the population ratio $S_{M}$
by (see Eq. \ref{e:MB}) \begin{equation}
C_{M}=\left(\sigma_{\star}/\sigma_{M}\right)^{3}S_{M}\,.\label{e:CMSM}\end{equation}
Specifically, if the different unbound mass groups are in equipartition
(as assumed by BW77), then $C_{M}\!=(M/\Ms)^{3/2}S_{M}$. Conversely,
if the unbound stellar population has achieved equilibrium by violent
relaxation \citep{lyn67}, then all mass groups are expected to have
similar velocity dispersions, and $C_{M}\!=\! S_{M}\!=\! S_{\star}$.
Here we assume violent relaxation boundary conditions%
\footnote{Eqs. (\ref{e:BCs}, \ref{e:CMSM}) generalize the definition of $g_{M}(x)$
and correct an error in the relation between $C_{M}$ and the population
ratio, which appeared in \citet{hop+06b}. The density profile presented
in \citet{hop+06b} closely approximates the violent relaxation solution
(cf Fig. \ref{f:FPGCmodel} here), and not, as stated there erroneously,
the isothermal solution.%
}, since these are suggested by the lack of an observed correlation
between scale height and spectral type of old stars in the Milky Way
\citep{kro+93},%
{} by the absence of an observed vertical color gradient in edge-on
disk galaxies \citep{deG+00}, and by the weak observed spatial color
gradients in elliptical galaxies \citep[e.g. ][]{bor+83,coh86}. %
{} We confirm the conclusion of BW77 that the steady state DFs do not
depend strongly on these specific choices of boundary conditions (cf
comparison with BW77 equipartition models in Fig. \ref{f:pLH_VR}).

The spatial density profile that corresponds to the DF is \begin{equation}
n_{M}(r)=\frac{2}{\sqrt{\pi}}\nstr\int_{-\infty}^{r_{\star}/r}g_{M}(x)\sqrt{\rstr/r-x}\mathrm{d}x\,.\label{e:nM}\end{equation}

\section{The relaxational coupling parameter }

\label{s:rlxparm}

As shown below in \S \ref{s:results}, the nature of the mass segregation
solution, weak or strong, is determined by one parameter, which expresses
the relative strength of $M_{H}$--$M_{H}$ and $M_{H}$--$M_{L}$
interactions. These can be quantified by the corresponding diffusion
coefficients, $D(\Delta v_{\Vert})$, $D(\Delta v_{\Vert}^{2})$ and
$D(\Delta v_{\bot}^{2})$, which enter the Fokker-Planck equation
\citep[e.g.][Eq. 8-68]{bin+87}. The two quadratic scattering coefficients
are similar up to an order unity factor, and are approximated here
as equal, $D(\Delta v_{\Vert}^{2})\sim D(\Delta v_{\bot}^{2})\equiv D(\Delta v^{2})/2$.
In a 2-mass system we represent these coefficients for brevity by
the notation $D_{STp}$ where $S=L,H$ is the scattering star (light
or heavy), $T=L,H$ is the target star, and $p=s,f$ is the process
(scattering, $D(\Delta v^{2})$, or dynamical friction, $D(\Delta v_{\Vert})$).
There are 8 possible permutations of the diffusion coefficients, ($\{L,H\}\!\times\!\{L,H\}\!\times\!\{s,f\}$),
however, since $D_{STs}\!\propto\!2N_{S}M_{S}^{2}$ and $D_{STf}\!\propto\! N_{S}M_{S}(M_{S}+M_{T})$,
there are only 4 distinct combinations of the mass and number ratios%
\footnote{The quadratic scattering terms do not depend on the mass of the scattered
target star because acceleration by a given gravitational force is
equal for all masses. The quadratic terms express a random walk in
velocity due to interactions with field stars on a timescale of $T_{R}\T{\sim}v^{2}/D(\Delta v^{2})$.
In contrast, the first order dynamical friction term expresses the
back-reaction of the field stars to the target star, and therefore
does depend on its mass. The first-order term expresses a systematic
deceleration on a timescale of $T_{\mathrm{df}}\T{\sim}v/D(\Delta v_{\Vert})$.%
}. 

The relaxational coupling parameter $\Delta$ describes the competition
between the self-coupling of the heavy stars and the light-heavy coupling
in terms of global properties of the system, the mass and number ratios,\begin{eqnarray}
\Delta & \equiv & \frac{D_{HHs}+D_{HHf}}{D_{LHs}+D_{LHf}}\simeq\frac{N_{H}M_{H}^{2}}{N_{L}M_{L}^{2}}\times\frac{4}{3+M_{H}/M_{L}}\,.\label{e:delta}\end{eqnarray}
The definition of $\Delta$ can in principle be generalized to a multi-mass
stellar population by specifying the light/heavy mass boundary, $M_{L/H}$
and performing a weighted integration over the mass function%
\footnote{E.g. $\Delta\!=\!\left.4\left\langle NM^{2}\right\rangle _{H}\right/\left(3\left\langle NM^{2}\right\rangle _{L}\!+\!\left\langle M_{H}\right\rangle \left\langle NM\right\rangle _{L}\right)$,
where for a normalized PMF $\mathrm{d}N/\mathrm{d}M$, $\left\langle NM^{k}\right\rangle _{L,H}\!=\!\int_{M_{L},M_{L/H}}^{M_{L/H},M_{H}}M^{k}\left(\mathrm{d}N/\mathrm{d}M\right)\mathrm{d}M$,
and $\left\langle M_{H}\right\rangle \!=\!\left.\int_{M_{L/H}}^{M_{H}}M\left(\mathrm{d}N/\mathrm{d}M\right)\mathrm{d}M\right/\int_{M_{L/H}}^{M_{H}}\left(\mathrm{d}N/\mathrm{d}M\right)\mathrm{d}M$.%
}. However, as shown in \S \ref{s:PMF} below, an evolved stellar
population (coeval or continuously star-forming) such as is expected
near a MBH, is well approximated by a 2-mass system for this purpose.

\subsection{Asymptotic mass segregation limits}

\label{ss:limits}

\emph{The weak segregation limit (the Bahcall-Wolf solution).} In
the limit $\Delta\!\rightarrow\!\infty$, which is the zero-flow ($Q_{M}\T{\rightarrow}0$)
limit (\S \ref{s:results}), the heavy stars dominate the population
and relax to the single mass cusp $\alpha_{H}\!=\!7/4$ ($p_{H}\!=\!1/4$).
The light stars heat by scattering against the effectively infinite
reservoir of heavy stars and diffuse to higher energies, thereby settling
to a flatter cusp with $\alpha_{L}\!\rightarrow\!3/2$ ($p_{L}\!\rightarrow\!0$).

\emph{The strong segregation limit.} In the limit  $\Delta\!\rightarrow\!0$
and when $M_{H}\T{\gg}M_{L}$, the light stars behave as a single
mass population with $\alpha_{L}=7/4$ ($p_{L}\!=\!1/4$). The rare
heavy stars sink to the center by dynamical friction against the effectively
infinite reservoir of light stars. Their steady state distribution
can be simply derived by noting that the dynamical friction force
is approximately $F\propto\rho_{L}/v^{2}\propto r^{1-\alpha_{L}}$,
where $v\!\propto\! r^{-1/2}$ is the Keplerian velocity \citep{cha43}.
The torque on the heavy star (assuming a circular Keplerian orbit,
$L\!\propto\! r^{1/2}$) is then $\dot{L}\!\propto\! Fr\!\propto\! r^{2-\alpha_{L}}\!\propto\! r^{-1/2}\dot{r}$,
so that $\dot{r}\!\propto\! r^{5/2-\alpha_{L}}$. In steady state,
the heavy star current, $\dot{N}_{H}$, is independent of radius,
$\dot{N}_{H}\!\propto\!\dot{r}r^{2}n_{H}(r)\!=\!\mathrm{const}$,
so that the number density of the heavy stars must scale as $n_{H}\!\propto\! r^{-(9/2-\alpha_{L})}\!\propto\! r^{-\alpha_{H}}$.
It then follows that \begin{equation}
\alpha_{H}=9/2-\alpha_{L}=11/4\qquad(p_{H}=5/4)\,.\end{equation}
Equivalently, this result can be obtained by expressing the Fokker-Planck
equations explicitly in terms of the diffusion coefficients as function
of energy \citep[Eqs. 51a,b]{lig+77}, and solving them under the
assumption that only the drift term (dynamical friction) contributes.
In practice, it is unlikely that physical systems actually have $\Delta\T{\ll}0.1$
(\S \ref{s:results}); $\alpha_{H}\T{=}11/4$ should be considered
an upper limit on the logarithmic slope of the number density distribution
of the heavy stars.

\section{The present day mass function}

\label{s:PMF}

\begin{figure}
\noindent \begin{centering}
\includegraphics[width=1\columnwidth]{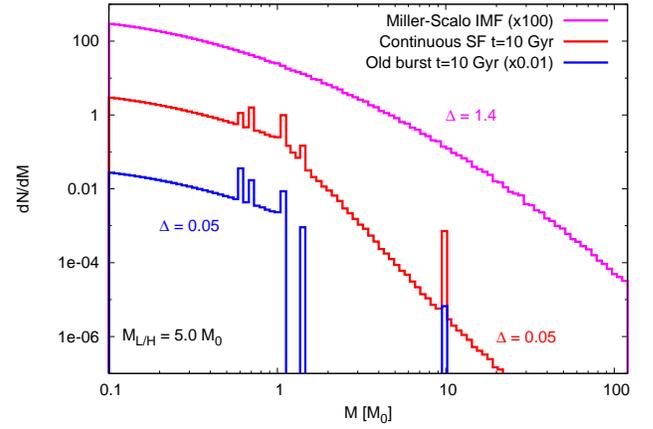}
\par\end{centering}

\caption{\label{f:PMFdelta}The PMF and the relaxation coupling parameter for
the Miller-Scalo IMF \citep{mil+79} (top line) assuming 10 Gyr of
continuous star formation (middle line) or a 10 Gyr old coeval population
(bottom line). The IMF was evolved using the Solar metallicity stellar
tracks of \citet{sch+92a} and the zero-age main sequence progenitor
masses were mapped to compact remnants by $ $ $0.8\!<\! M/M_{\odot}\!<\!1.5\!\rightarrow\!0.6\, M_{\odot}$
WD, $1.5\!<\! M/M_{\odot}\!<\!2.5\!\rightarrow\!0.7\, M_{\odot}$
WD, $2.5\!<\! M/M_{\odot}\!<\!8.0\!\rightarrow\!1.1\, M_{\odot}$
WD, $8.0\!<\! M/M_{\odot}\!<\!30.0\!\rightarrow\!1.4\, M_{\odot}$
NS, $30.0\!<\!\Ms/M_{\odot}\!\rightarrow\!10\, M_{\odot}$ SBH \citep{ale05}.
$\Delta$ was estimated for a light/heavy mass boundary $M_{L/H}\!=\!5\,\Mo$
(the exact choice of $M_{L/H}$ does not affect the value of $\Delta$
for the evolved PMFs). \protect \\
}

\end{figure}

The value of $\Delta$ (Eq. \ref{e:delta}) depends on the population's
PMF. So-called universal IMFs, which extend all the way from the brown
dwarf boundary $M_{1}\!\sim\!0.1\,\Mo$ to $M_{2}\!\gtrsim\!100\,\Mo$
(e.g. the \citealt{sal55} IMF and its subsequent refinements, the
\citealt{mil+79} and \citealt{kro01} IMFs), result in evolved populations
(old coeval star-bursts or continuously star forming populations)
that naturally separate into two mass scales, the $\sim\!1\,\Mo$
scale of low-mass main-sequence dwarfs, white dwarfs and neutron stars,
and the $\sim\!10\,\Mo$ scale of stellar black holes, and typically
have $\Delta\!<\!0.1$ (Fig. \ref{f:PMFdelta}). Such evolved populations
are well-approximated by the simple 2-mass population model. 

Generally, 10 Gyr old, continuously star-forming populations with
a power-law IMF, $\mathrm{d}N/\mathrm{d}M\!\propto\! M^{-\gamma}$,
have $\Delta\!<\!1$ for $\gamma\!\gtrsim\!1.8$, and $\Delta\!>\!1$
for $\gamma\!\lesssim\!1.8$ (assuming $M_{L}\!=\!0.1\,\Mo$, $M_{H}\!=\!120\,\Mo$,
$M_{L/H}\!=\!5\,\Mo$). This implies that even flatter IMFs than the
Salpeter IMF ($\gamma\!=\!2.35$) lead to strong mass segregation. 

It is not clear what is the typical PMF, and the corresponding value
of $\Delta$, in galactic nuclei. There are indications that star
formation deep in the potential well of a MBH can be very different
from that in the field \citep{lev+03,mil+04,pau+06b,lev07}. For example,
the volume-averaged stellar population in the central few pc of the
GC is reasonably well approximated by a 10 Gyr old, continuously star-forming
population with a universal IMF \citep{ale+99a} and has $\Delta\!<\!0.1$.
In contrast, a recent analysis of late type giants in the inner $\sim\!1$
pc of the GC suggests that the IMF there could be a flat $\gamma\!\sim\!0.85$
power-law \citep{man+07}. This corresponds to $\Delta\!>\!10$ in
the inner $\sim\!1$ pc, if that region evolved as an isolated system.
However, that is unlikely, since the PMF reflects not only the local
star formation, but also the volume-averaged population in the inner
few pc, which contributes unbound stars and stars that diffuse in
phase space from lower to higher energies. It is therefore quite possible
that in some galactic nuclei the weak mass segregation solution is
realized, while in others the strong mass segregation, depending on
the system-to-system scatter in the PMF.

\section{Results}

\label{s:results}

\begin{figure}
\noindent \begin{centering}
\includegraphics[width=1\columnwidth]{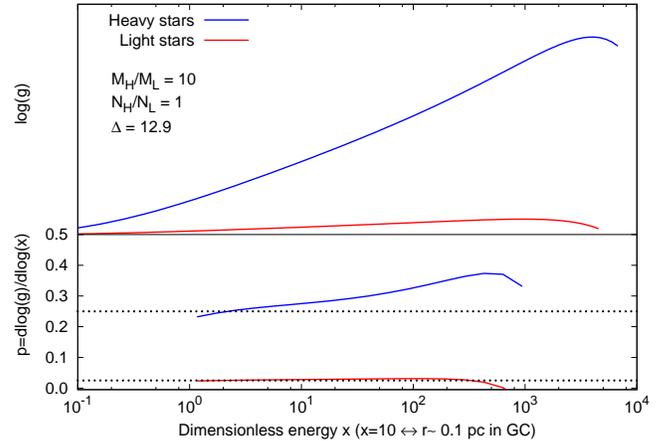}
\par\end{centering}

\caption{\label{f:FPsol} The steady state DF (on an arbitrary logarithmic
scale) and its local logarithmic slope away from the boundaries, for
$M_{H}/M_{L}\!=\!10$ and $ $ $N_{H}/N_{L}\!=\!1$ ($\Delta\!=\!12.9$).
The loss-cone term is neglected. The boundary conditions are a sink
at $x_{D}=10^{4}$ and violently relaxed (equal velocity dispersion)
unbound stars. The asymptotic BW77 solution is indicated by the dotted
lines ($p_{L}\!=\!1/40$, $p_{H}\!=\!7/4$).}

\end{figure}

\begin{figure*}
\noindent \begin{centering}
\begin{tabular}{cc}
\includegraphics[width=1\columnwidth]{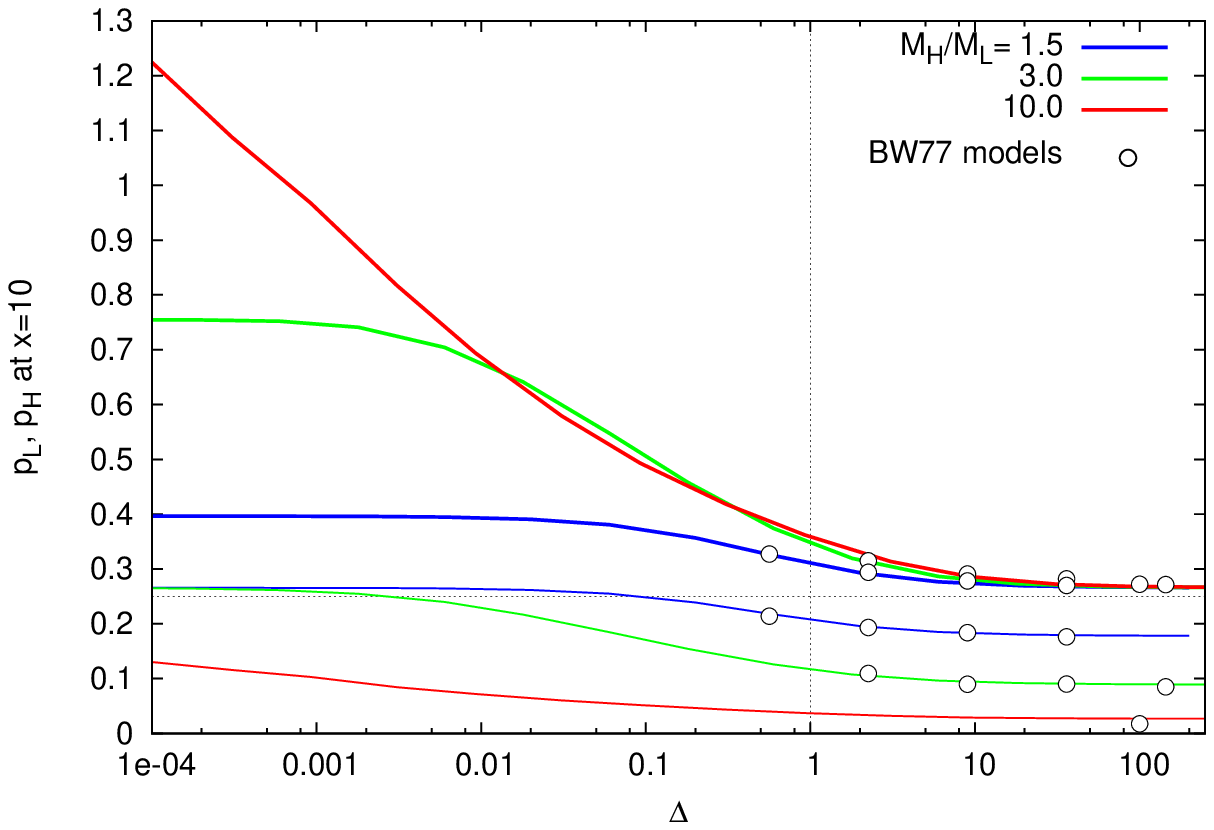} & \includegraphics[width=1\columnwidth]{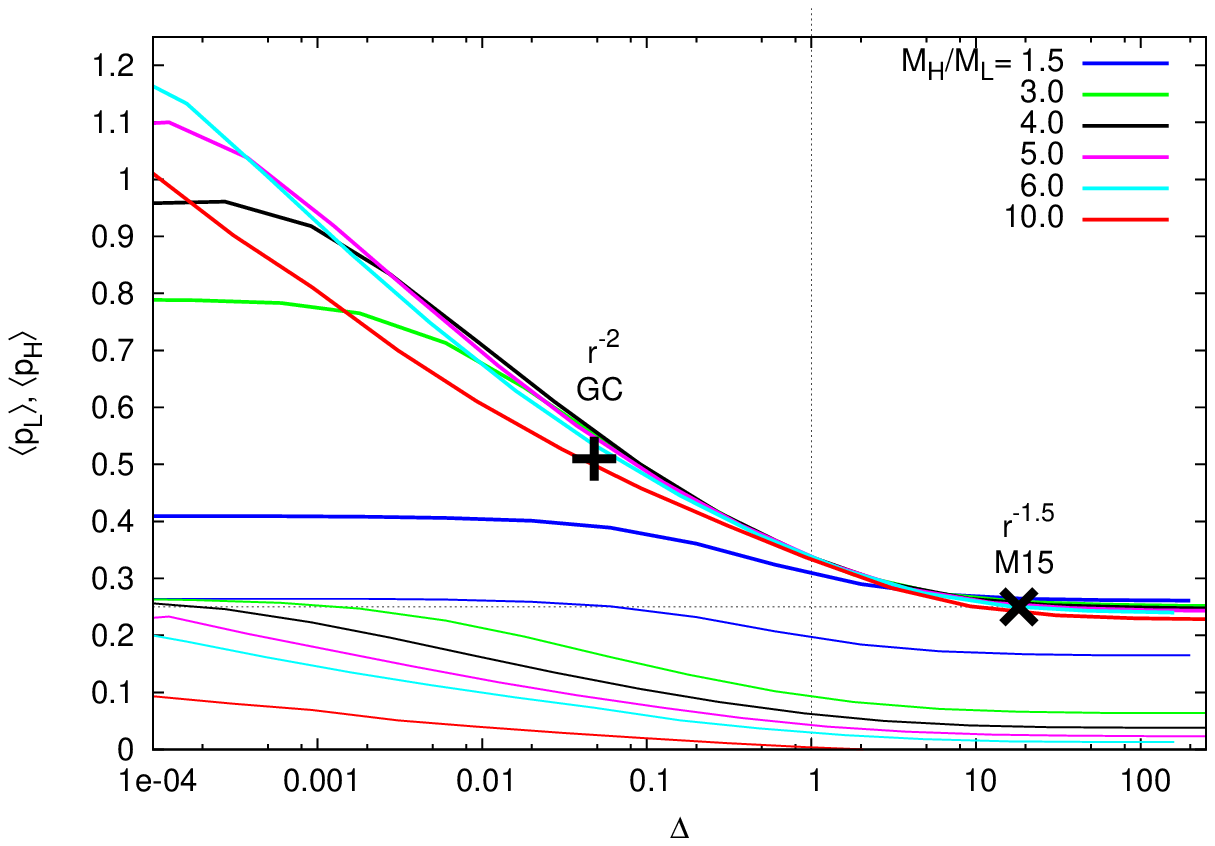} \tabularnewline
\end{tabular}
\par\end{centering}

\caption{\label{f:pLH_VR} The power-law indices $p_{L,H}$ as functions of
$\Delta$ for various mass ratios, assuming violent relaxation boundary
conditions, and neglecting the loss-cone term. Left: The indices,
as derived from the logarithmic slopes of $g_{L,H}$ at $x\!=\!10$,
compared to the logarithmic slopes of the DFs calculated by \citet{bah+77}.
Right: The indices, as derived from the average logarithmic slopes
of the densities $3/2+p_{L,H}$ in the range $r\!=\!0.003$--$0.3$
pc (for $\rstr\!=\!2$ pc) with the expected $p_{H}$ for a model
of the GC assuming continuous star formation with the Miller-Scalo
IMF \citep[Fig. \ref{f:PMFdelta};][]{ale+99a,ale05}, and for a mass-function
model of globular cluster M15 \citep{mur+97}, assuming it harbors
an IMBH.}

\end{figure*}

Figure (\ref{f:FPsol}) shows an example of a steady state solution
of the DFs calculated for a 2-mass system in the weak segregation
regime with $M_{L}/M_{H}\!=\!10$ and $N_{H}/N_{L}\!=\!1$ ($\Delta\!\simeq\!13$),
neglecting the loss-cone term. While the logarithmic slopes of the
DFs are not constant, they vary only slightly away from the boundaries
at $x=0$ and $x_{D}$. 

When comparing the DF with observations of stars in the GC, a relevant
energy scale is $x\!\sim\!10$, which corresponds to orbits with a
semi-major axis of $a\!=\!\rstr/2x=0.1$ pc for $\rstr\!=\!2$ pc
in the GC. On that scale the cusp is Keplerian, but there are still
enough observed stars for meaningful statistics \citep[e.g.][]{sch+07}.
Figure (\ref{f:pLH_VR} L) shows the local logarithmic slopes of the
DFs, $p_{L}$ and $p_{H}$, at $x\!=\!10$, as function of $\Delta$,
for the same mass ratios modeled by BW77, $M_{H}/M_{L}\!=\!1.5$,
$3$ and $10$. The transition between the weak and strong mass segregation
solutions at $\Delta\T{\sim}1$ is clearly seen. It is interesting
to note that the set of models calculated by BW77 all happen to lie
at $\Delta\!\gtrsim\!1$, which explains why they found that {}``\ldots{}there
are no dramatic changes in the shapes of the curves, despite the fact
that the unbound population ratio was varied by a factor of 64.'',
and wrongly concluded that the weak segregation solution is a universal
solution.

A useful average of the logarithmic slope is obtained from the corresponding
stellar density curve (Eq. \ref{e:nM}, cf Fig. \ref{f:FPGCmodel}),
which can be directly related to the observable stellar density distribution.
A comparison of Figs. (\ref{f:pLH_VR} L) and (\ref{f:pLH_VR} R)
shows that the shape of the curves is hardly affected by the method
chosen for deriving the logarithmic slopes. We also repeated these
calculations with a loss-cone term (Eq. \ref{e:FPeq}), and confirmed
the BW77 conclusion that the effect on the shape of the DF curve is
small, thereby justifying its omission. 

The asymptotic limits (\S \ref{ss:limits}) are also clearly seen
in Fig. (\ref{f:pLH_VR}). When $\Delta\!\rightarrow\!\infty$, the
weak segregation BW77 solution holds, with $p_{H}\!=\!0.25$ and $p_{L}=M_{L}/4M_{H}$.
When $\Delta\!\rightarrow\!0$, the light stars, which dominate the
population, assume the single mass population DF, $p_{L}\!=\!1/4$,
while the heavy stars concentrate to the center. For low mass ratios,
$M_{H}/M_{L}\!\lesssim\!4$, where dynamical friction is less efficient,
the heavy stars approximately obey the BW77 relation, $p_{H}\!=\!(M_{H}/M_{L})p_{L}\!=\! M_{H}/4M_{L}$.
For higher mass ratios, $M_{H}/M_{L}\!\gtrsim\!4$, the heavy stars
approach the dynamical friction limit, $p_{H}\rightarrow5/4$. 

\begin{figure}
\includegraphics[width=1\columnwidth]{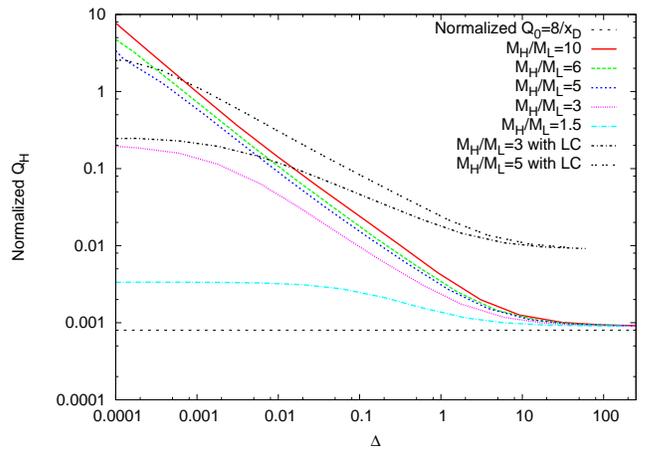}

\caption{\label{f:Delta_flow}The normalized current of the heavy stars, $\widehat{Q}_{H}$
(Eq. \ref{e:Qnorm}) , as function of the relaxational coupling parameter,
$\Delta$. The horizontal line marks $\widehat{Q}_{0}\T{=}8/x_{D}$,
the zero-flow in a single mass population \citep{bah+76}. For $\Delta\T{\gg}1$,
$\widehat{Q}_{H}\T{\sim}\widehat{Q}_{0}\T{\sim}0$. However, for $\Delta\T{\ll}1$
in the strong mass-segregation regime, $\widehat{Q}_{H}\T{\gtrsim}O(1)$,
indicating that massive stars flow inwards at the maximal possible
rate, the dynamical friction rate. The transition between the two
flow regimes is also apparent in the energy-averaged currents in models
with a loss-cone term.}

\end{figure}

The transition between the weak and strong mass segregation solutions
is a reflection of the breakdown of the zero-flow assumption as $\Delta\T{\rightarrow}0$.
The dimensional scale $I_{\star}$ of the dimensionless stellar current
into the MBH, $Q_{M}(\epsilon)\T{=}I_{M}(\epsilon)/I_{\star}$, is
$I_{\star}\T{\sim}O(N_{\Ms}(<\! r_{\star})/t_{\star})$  (note that
when the loss-cone is neglected, the steady-state current is independent
of energy). This is also the order of magnitude of the stellar current
into the MBH in a single mass population out of equilibrium. However,
\citet[Eq. 63]{bah+76} show that, neglecting the loss-cone, the steady-state
current in a single mass population is $Q_{M}\T{\sim}8/x_{D}\sim0$,
and that steady state zero-flows further imply that $p_{H}(\varepsilon)/M_{H}=p_{L}(\varepsilon)/M_{L}$
in multi-mass populations \citep[Eqs. 41-46]{bah+77}. The relevant
physical timescale for the heavy star current $Q_{H}$ is not the
$M_{L}$--$M_{L}$ interaction timescale $1/t_{\star}\T{=}\Gamma_{L}\T{\propto}D_{LLs}+D_{LLf}\sim4N_{L}M_{L}^{2}$
(here we choose $\Ms\T{=}M_{L}$), but rather the rate of $M_{H}$--$M_{H}$
and $M_{L}$--$M_{H}$ interactions, $\Gamma_{H}\T{\propto}D_{HHs}+D_{LHs}+D_{HHf}+D_{LHf}\sim N_{L}M_{L}^{2}\left[4+M_{H}/M_{L}+4N_{H}M_{H}^{2}/N_{L}M_{L}^{2}\right]$.
To compare meaningfully the heavy star current $Q_{H}$ across the
range of our models, we rescale it to $1/\Gamma_{H}$ instead of $t_{\star}$,
and normalize by the number of heavy stars to obtain the current per
star, \begin{equation}
\widehat{Q}_{H}\T{=}Q_{H}\frac{N_{L}}{N_{H}}\frac{\Gamma_{L}}{\Gamma_{H}}\T{=}Q_{H}4\frac{N_{L}}{N_{H}}\left/\left(4+\frac{M_{H}}{M_{L}}+4\frac{N_{H}M_{H}^{2}}{N_{L}M_{L}^{2}}\right)\right.\,.\label{e:Qnorm}\end{equation}
 Figure (\ref{f:Delta_flow}) shows $\widehat{Q}_{H}$ as function
of $\Delta$ for different mass ratios. For $\Delta\T{>}1$, which
is the range explored by BW77, we confirm their zero-flow result.
In contrast, for $\Delta\T{\ll}1$, we find that the zero-flow assumption
no longer holds, $\widehat{Q}_{H}\T{\gtrsim}O(1)$, and the heavy
stars sink to the MBH at the maximal possible rate, the dynamical
friction rate. Figure (\ref{f:Delta_flow}) also shows that the physically
required inclusion of a loss-cone term in the Fokker-Planck equations
(Eq. \ref{e:Rm}) increases the current, but the energy averaged flow
in the weak segregation limit $\Delta\T{\gg}1$, while not quite as
small as without a loss-cone, is still strongly suppressed%
\footnote{In the presence of a loss-cone, the current in the weak segregation
limit is set by the diffusion rate into the loss-cone at $x_{\mathrm{diff}}$
(\S\ref{s:FPeq}), $I_{H}\T{\sim}N_{H}(x_{\mathrm{diff}})/[t_{H}\log([J_{c}(x_{\mathrm{diff}})/J_{lc})]$,
where $t_{H}\T{\sim}\Gamma_{H}^{-1}$ is the ($M_{H}$-dominated)
relaxation time (cf Eq. \ref{e:Rm}). It then follows that $\widehat{Q}_{H}(x_{\mathrm{diff}})\T{\sim}I_{H}(x_{\mathrm{diff}})/(N_{H}\Gamma_{H})\T{\sim}[1/(2x_{\mathrm{diff}})]^{3-\alpha_{H}}/\log[(c/\sigma_{\star})/\sqrt{32x_{\mathrm{diff}}}]\T{\sim}0.01$
for $x_{\mathrm{diff}}\T{=}10$.%
}. It then follows that the relation $p_{H}/M_{H}\T{\sim}p_{L}/M_{L}$
is approximately satisfied in this limit even with a loss-cone, as
found by BW77 and confirmed here. In the limit strong segregation
limit $\Delta\T{\ll}1$, the dynamical friction-driven current of
the heavy stars is insensitive to the presence or absence of a loss-cone,
since $T_{\mathrm{df}}\T{\sim}(M_{L}/M_{H})T_{R}$.

Figure (\ref{f:pLH_VR} R) shows the expected segregation of the massive
stellar objects in the GC (based on the old, continuous star-formation
PMF model, Fig. \ref{f:PMFdelta}), and in globular cluster M15, assuming
it harbors an IMBH (based on the tentative PMF model of \citealt{mur+97},
which assumes that the most massive remnants in the cluster are $1.4\,\Mo$
neutron stars). The GC is expected to lie in the strong relaxation
regime ($\Delta\!=\!0.05$), while M15 in the weak relaxation regime
($\Delta\!\sim\!18$).

We refrain here from a more detailed quantitative analysis of the
numerical results because the exact values of $p_{L,H}(\Delta)$ depend
somewhat on the way these are evaluated (locally, or from the density
curve), and because the convergence of our numerical Fokker-Planck
solver becomes progressively worse in the limits $\Delta\!\rightarrow\!0$
and $M_{H}/M_{L}\!\gg\!1$. However, any numerical inaccuracies in
those limits are unlikely to have practical implications, since realistic
stellar systems are not expected to have $\Delta\!\lesssim\!0.01$
(a possible exception could be the hypothesized steady flow of IMBHs
from dense stellar clusters to the MBH, \citealt{por+06}).

\section{Discussion and summary}

\label{s:discuss}

\subsection{Strong segregation and other relaxation processes}

Strong mass segregation, or mass segregation instability, shares some
common features with the Spitzer, or equipartition, instability in
a cluster \citep{spi69}, where the heavy stars decouple from the
light ones and evolve away from equipartition by forming a dense sub-system
in the center with a much higher velocity dispersion, $\sigma_{H}^{2}/\sigma_{L}^{2}\T{>}M_{L}/M_{H}$.
However, these two instabilities are distinct effects. The Spitzer
instability occurs only when the heavy stars are relatively common
in the population, $N_{H}M_{H}^{5/2}/N_{L}M_{L}^{5/2}>\beta$, where
$\beta\!\simeq0.16$ (for $M_{H}\!\gg\! M_{L}$ and $N_{H}M_{H}\!\ll\! N_{L}M_{L}$).
In contrast, in the Keplerian potential near a MBH, the Jeans equation
dictates that $\sigma_{H}^{2}/\sigma_{L}^{2}\T{=}(5/2+p_{L})/(5/2+p_{H})\T{\sim}1$
\citep{ale+01a}, and so equipartition is never achieved, irrespective
of the heavy-to-light mass or number ratios. Strong mass segregation
is an instability in the spatial distribution of the heavy stars,
which occurs in the opposite limit to the Spitzer instability, when
the heavy stars are relatively rare in the population, $\Delta\T{\sim}N_{H}M_{H}/N_{L}M_{H}\T{<}1$
(for $M_{H}\T{\gg}M_{L}$, Eq. \ref{e:delta}). 

\citet{per+07} focused on the relaxation of light objects, stars,
by heavy objects, massive perturbers (e.g. giant molecular clouds
or clusters, with $M_{H}\T{\gg}M_{L}$), and expressed the efficiency
of massive perturber-induced relaxation relative to star-star relaxation
by the parameter $\mu_{2}\T{=}N_{H}M_{H}^{2}/N_{L}M_{L}^{2}$ ($\sim\!(D_{HLs}+D_{HLf})/(D_{LLs}+D_{LLf})\T{\simeq}3N_{H}M_{H}^{2}/4N_{L}M_{L}^{2}$).
When $\mu_{2}\T{\gg}1$, massive perturbers dominate stellar relaxation.
Here we focus on the dynamics of the heavy objects, the SBHs, and
so the relaxational coupling parameter $\Delta$ is defined to also
take into account the interactions between the heavy masses. The parameter
$\mu_{2}$ addresses the question ``Which mass component dominates
the relaxation of the light stars?'', while $\Delta$ addresses the
question ``Which mass component dominates the dynamics of the heavy
stars?''. The two are related by $\Delta\T{=}\mu_{2}4/(3+M_{H}/M_{L})$.

Our approximate treatment of the mass segregation process neglects
the relaxation of angular momentum to near radial ({}``loss-cone'')
orbits. This, rather than diffusion in energy, is the primary channel
for stellar destruction by the MBH. A full treatment of the mass segregation
problem in ($E,J$) phase space \citep[e.g.][]{coh+78} is beyond
the scope of this work. Here we follow BW77, who treated the problem
approximately in $E$ only, and who further showed that the neglect
of an effective loss-cone term (Eq. \ref{e:FPeq}) did not much change
the shape of the DFs. We confirmed that this conclusion also holds
for our mass segregation models (Fig. \ref{f:Delta_flow}).

\subsection{Possible implications of strong mass segregation}

Strong mass segregation occurs in stellar systems with a relatively\emph{
lower} fraction of SBHs, that reach a \emph{higher} central concentration
of SBHs very close to the MBH, compared to systems with a higher fraction
of SBHs that undergo weak segregation. To compare two such systems,
and to determine which has more SBHs enclosed inside some given volume
around the MBH, it is necessary to specify the comparison procedure
(e.g. assuming an equal total stellar mass; or the same MBH mass and
the $\Mbh/\sigma$ relation; or an equal number of SBHs within $r_{h}$).
The choice depends on the question of interest. Here we do not address
such quantitative issues, but limit ourselves to briefly listing some
processes that are expected to be affected by the degree of segregation.

\emph{Accelerated relaxation. }The degree of mass segregation affects
both the non-coherent 2-body relaxation timescale, $ $which scales
as $1\left/\int M^{2}(\mathrm{d}N/\mathrm{d}M)\mathrm{d}M\right.$
(see \S \ref{s:intro}), and the resonant relaxation timescale, which
scales as as $\int M(\mathrm{d}N/\mathrm{d}M)\mathrm{d}M\left/\int M^{2}(\mathrm{d}N/\mathrm{d}M)\mathrm{d}M\right.$
\citep{rau+96}. In particular, the stronger the mass segregation,
the shorter is the resonant relaxation timescale, which does not depend
on the number of stars, but only on their typical mass. Efficient
resonant relaxation near the MBH may affect stellar orbits and accretion
disk dynamics there (see below). 

\emph{GW event rates.} The GW EMRI rate is determined by the number
of potential GW sources within the critical radius $r_{\mathrm{crit}}$,
which demarcates the boundary between compact object that inspiral
into the MBH those that plunge (infall) into it. The critical radius
is a function of the relaxation time, and to good approximation the
EMRI rate is $\Gamma\T{\sim}N[r_{crit}(T_{R})]/T_{R}$ \citep{hop+05}.
Strong segregation will affect the EMRI rate both by modifying the
2-body relaxation time and by affecting the number of stars enclosed
inside $r_{\mathrm{crit}}$, as well as by decreasing the resonant
relaxation timescale. Similarly, the rates of detectable GW bursts
from fly-bys near the Galactic MBH strongly depend on the number of
SBHs near it \citep{hop+07}. Strong segregation may also affect GW
emission from close SBH--SBH interactions in a very dense cusp \citep{ole+08b}.

\emph{SBH--star interactions.} A higher central concentration of SBHs
affects the probability of SBH--star interactions, which can lead
to the randomization of stellar orbits, the heating of a stellar disk
\citep{per+08}, the 3-body exchange capture of massive young stars
near the MBH \citep{ale+04}, or the ejection of hyper-velocity stars
\citep{ole+08}.

\emph{SBH--accretion disk interactions.} A higher central concentration
of SBHs within $10^{1-3}$ gravitational radii of the MBH could exert
coherent torques on the accretion disk, warp it and possibly affect
its hydrodynamics (Bregman \& Alexander, 2008, in prep.). The SBHs
may shock the disk as they cross it, experience drag by it, and be
carried by it to the MBH \citep[e.g.][]{art+93,nay+04,sub+05}.

\emph{Enhanced gravitational lensing. }SBHs projected near the Einstein
angle of the MBH can strongly modify the gravitational lensing properties
of the MBH,\emph{ }in a way similar to the effect of a planet orbiting
a Galactic star that is lensing a background source \citep{ale+01c,cha+01a}.

\subsection{Summary}

We show that the steady state solution of a relaxed multi-mass stellar
system around a MBH has two branches: the known weak (Bahcall-Wolf)
mass segregation solution, where the difference in the degree of central
concentration of the light and heavy stars is relatively small, and
a newly discovered strong segregation solution, where the difference
is much larger. The nature of the solution is determined by the global
properties of the system (the mass ratio between the heavy and light
stars, $M_{H}/M_{L}$, and their number ratio far from the MBH, $N_{H}/N_{L}$)
through the relaxational coupling parameter, $\Delta\!=\!4N_{H}M_{H}^{2}\left/\left[N_{L}M_{L}^{2}(3\!+\! M_{H}/M_{L})\right]\right.$.
Strong mass segregation occurs when the heavy stars are relatively
rare in the population $(\Delta\T{\ll}1)$, and sink to the center
by dynamical friction. Weak mass segregation occurs when the heavy
stars are common in the population ($\Delta\T{\gg}1$) and settle
to the single mass stellar cusp solution. We show that relaxed old
coeval or continuously star-forming populations with a universal IMF
typically have $\Delta\T{<}0.1$, and thus settle to the strong mass
segregation solution around a MBH.

\acknowledgements{We are grateful to S. Tremaine and M. Freitag for useful discussions.
TA is supported by ISF grant 928/06, ERC Starting Grant 202996 %
{}and a New Faculty grant by Sir H. Djangoly, CBE, of London, UK. CH
is supported by a Veni scholarship from the Netherlands Organization
for Scientific Research (NWO).}

\bibliographystyle{apj}

\end{document}